\documentclass[twocolumn,floats,prl,amsmath,amssymb,superscriptaddress,showpacs]{revtex4}
\usepackage{epsfig}
\usepackage{graphicx}

\begin{document}
\title{Cloning the entanglement of a pair of quantum bits}

\author{Louis-Philippe Lamoureux}
\affiliation{Quantum Information and Communication, Ecole Polytechnique, CP 165,\\ Universit\'e Libre de Bruxelles, 1050
Brussels, Belgium}

\author{Patrick Navez}
\affiliation{Quantum Information and Communication, Ecole Polytechnique, CP 165,\\ Universit\'e Libre de Bruxelles, 1050
Brussels, Belgium}

\author{Jarom\'\i r Fiur\' a\v sek}
\affiliation{Quantum Information and Communication, Ecole Polytechnique, CP 165,\\ Universit\'e Libre de Bruxelles, 1050
Brussels, Belgium}
\affiliation{Department of Optics, Palack\'{y} University, 
17. listopadu 50, 77200 Olomouc, Czech Republic}

\author{Nicolas J. Cerf}
\affiliation{Quantum Information and Communication, Ecole Polytechnique, CP 165,\\ Universit\'e Libre de Bruxelles, 1050
Brussels, Belgium}

\begin{abstract}
It is shown that any quantum operation that perfectly clones the entanglement of
all maximally-entangled qubit pairs cannot preserve separability. This ``entanglement no-cloning'' principle naturally suggests that some approximate cloning of entanglement is nevertheless allowed by quantum mechanics. 
We investigate a separability-preserving optimal cloning machine that duplicates all maximally-entangled states of two qubits, resulting in 0.285 bits of entanglement per clone, while a local cloning machine only yields 0.060 bits of entanglement
per clone.
\end{abstract}

\maketitle

Quantum entanglement is known to be a resource that is central to many quantum information processes such as quantum teleportation, quantum cryptography, or quantum computing \cite{nielsen}. In view of this, much work has been devoted to defining measures of entanglement 
or to investigating the best information-theoretical use of entanglement.
Despite the fact that entanglement is a very fragile resource, extremely sensitive to decoherence, several techniques have been developed in order to overcome decoherence, namely quantum error correction or entanglement purification (see \cite{BDSW}). 
Out of these many studies of entanglement, none has so far addressed 
the issue of whether (and how well) entanglement can be cloned.
\par

There has been a vast literature about the question 
of cloning quantum states. First of all, the {\em no-cloning} theorem has 
been stated \cite{WZ}, which precludes the perfect copying of an 
arbitrary quantum state.
Then, imperfect quantum cloning machines (QCM) have been introduced, 
which duplicate an arbitrary qubit state with the highest possible 
fidelity \cite{BH}. Since then and more recently,
a large variety of QCMs have been devised, 
with the purpose of cloning equally well a given set of states in a space
of arbitrary dimension (see \cite{Werner}).
\par

In this paper, we raise the question of whether quantum entanglement 
itself can be cloned or not. In order to simplify our analysis, 
we restrict ourselves to qubit pairs (dimension $2\times 2$). 
We show that the requirement of perfectly cloning the entanglement carried by a qubit pair in an arbitrary maximally-entangled (ME) state is incompatible with the requirement that separable qubit pairs remain unentangled via cloning. Of course, if we
restrict ourselves to four orthogonal ME states such as the Bell states
\begin{equation}
|\Phi^{\pm}\rangle=(|00\rangle\pm|11\rangle)/\sqrt{2}, \;\;
|\Psi^{\pm}\rangle=(|01\rangle\pm|10\rangle)/\sqrt{2},
\end{equation} then we can
very well make a Bell measurement of the original pair and subsequently prepare
an arbitrary number of clones in the measured ME state. 
However, this procedure does not work properly on the linear combinations 
of Bell states that remain ME states. 
Even if the clones are not required to be close to the original ME 2-qubit state but solely to be entangled, it remains impossible to fully preserve entanglement.
\par

Here we show that it is nevertheless possible to clone part of the original entanglement, much in the same way quantum states can be cloned imperfectly.
We construct a QCM that is universal over the set of ME 2-qubit states 
and argue that it effects an optimal cloning of entanglement. For this purpose, we exploit the property that the set of ME 2-qubit states is isomorphic to the set of {\em real} states in 4 dimensions \cite{BDSW}, from which we construct an optimal symmetric cloner that is covariant under local unitaries.
This cloner maximizes the amount of entanglement left in the clones 
(about 0.285 ebit per clone), 
while product states remain unentangled. Finally, we consider asymmetric QCMs and
investigate how entanglement is distributed among the clones by these transformations.  
\par

{\it Entanglement no-cloning principle}: 
Entanglement cannot be cloned perfectly, that is, if a quantum operation can be found that perfectly duplicates the entanglement of all ME states, then it necessarily does not preserve separability
(some separable states become entangled after cloning).

{\it Proof:}
We restrict ourselves to 2 qubits and consider two orthogonal ME states, e.g., $|\Phi^{\pm}\rangle$.
Assume that the entanglement of these states is perfectly cloned, i.e., 
the output states of the clones
remain ME even though they may differ from the input state.
The most general cloning transformation $U$ preserving the entanglement of these two states can be written as
\begin{eqnarray}
|\Phi^{\pm}\rangle |A\rangle \stackrel{U}{\rightarrow}
|e_a^{\pm}\rangle |e_b^{\pm} \rangle |A^{\pm} \rangle \ ,
\end{eqnarray}
where $|A\rangle$ denotes the initial state of the ancilla and
the blank copy, while $|A^{\pm}\rangle$ is the ancilla state after cloning. Thus, the states of the two clones $|e_a^{\pm}\rangle$ and $|e_b^{\pm} \rangle$ are some ME states. Now, the linear combination
$|{\tilde \Phi}\rangle= (|\Phi^+ \rangle + i|\Phi^- \rangle)/\sqrt{2}= 
(e^{i\pi/4}|00\rangle+e^{-i\pi/4}|11\rangle)/\sqrt{2}$
is still a ME state. By linearity, the above transformation yields the following output state
\begin{eqnarray}
|{\tilde\Phi}\rangle |A\rangle \stackrel{U}{\rightarrow}
(|e_a^+\rangle |e_b^+ \rangle |A^+ \rangle
+ i |e_a^-\rangle |e_b^- \rangle |A^- \rangle)/\sqrt{2} \ .
\end{eqnarray}
In order to preserve the full entanglement within each clone,
a necessary condition is that either $|e_a^+\rangle=|e_a^-\rangle$ or $|e_b^+\rangle=|e_b^-\rangle$. However, in each of these two cases, 
at least one of the clones is left in a ME state that is independent of the input state (within the space spanned by $|\Phi^\pm\rangle $) 
regardless of it being separable or not. For example, in the first case,
if the input is the separable state obtained as the linear combination
$|s \rangle=(|\Phi^+\rangle + |\Phi^-\rangle)/\sqrt{2}=|00\rangle$,
then the transformation gives
\begin{eqnarray}
|s\rangle |A\rangle \stackrel{U}{\rightarrow}
|e_a\rangle(|e_b^+\rangle |A^+\rangle + |e_b^-\rangle|A^-\rangle)/\sqrt{2} \ .
\end{eqnarray}
Clearly, the separability is not preserved here since the first clone
is maximally entangled. We therefore conclude that
no perfect cloning of entanglement is possible. $\Box$
\par

As a consequence, only imperfect QCMs that approximately reproduce the entanglement while preserving separability can be implemented. In the 
rest of this paper, we will be interested in separability-preserving
QCMs that yield clones with the highest achievable
entanglement for all ME input states. As shown later on, 
finding these QCMs is strongly related to finding
transformations that clone optimally and equally well 
the set of 2-qubit ME states.
Consider an arbitrary 2-qubit pure state 
\begin{equation}
|\Phi \rangle= \sum^3_{i=0} n_i|e_i \rangle
\end{equation}
written in the orthonormal basis made of the Bell states with particular phases (sometimes referred to as the magic basis \cite{BDSW}):
\begin{eqnarray}
|e_0\rangle=|\Phi^+\rangle, \; |e_1\rangle=i|\Phi^-\rangle, \;
|e_2\rangle=i|\Psi^+\rangle, \; |e_3\rangle=|\Psi^-\rangle,
\end{eqnarray}
where the amplitudes $n_i$ are normalized as $\sum_{i=0}^3 |n_i|^2=1$.
In this basis, the entanglement of formation $E$ of the state $|\Phi \rangle$ can be expressed in a very simple way as
\begin{equation}
E({\cal C}(\Phi))=H\bigg(\frac{1}{2}+\frac{1}{2}\sqrt{1-{\cal C}(\Phi)^2}\bigg) 
\ ,
\end{equation}
where $H$ is the binary entropy function and
\begin{equation}
{\cal C}(\Phi)=|\sum_i n_i^2|
\end{equation}
is called the concurrence \cite{BDSW,HW}.
Clearly, any {\em real} linear combination (up to an irrelevant global phase) of the 
magic basis elements is a ME state since ${\cal C}$ (and therefore $E$) is then equal to one.
Furthermore, {\em every} ME state can be expressed as a real linear combination of the magic basis elements.
For this reason, the problem of cloning the set of ME 2-qubit states boils down to constructing a transformation that optimally clones all real 4-dimensional states in the magic basis.
\par

This particular transformation can be found by following a method inspired
from \cite{Cerf}. We assume that the qubit pair to be cloned
is itself maximally entangled with a reference system (another qubit pair). 
We then explicitely construct the most general joint state describing the reference $\cal R$, the output clones $a$ and $b$, and an ancilla $\cal A$
after the cloning transformation
\begin{equation}\label{Psidefinition}
| \mathcal{S} \rangle_{{\cal R},a,b,{\cal A}} = \sum^3_{i,j,k,l=0} s_{ijkl}\; 
|l\rangle_{\cal R} |i\rangle_a|j\rangle_b|k\rangle_{\cal A} \ .
\end{equation}
(The reference, the two clones, and the ancilla are all 4-dimensional systems here.)
This state serves to completely define the cloning transformation:
the result of cloning the ME state $|\Phi\rangle=\sum_i n_i|e_i\rangle $ (with real $n_i$'s) is obtained by projecting ${\cal R}$ onto the complex conjugate $|\Phi^*\rangle$ (which is equal to $|\Phi\rangle$ here) \cite{Cerf}. Thus, the most general cloning transformation is defined as
\begin{equation} \label{cloningtrans}
|\Phi \rangle \to \sum^3_{i,j,k,l=0} s_{ijkl}\; n_l\; |i\rangle_a|j\rangle_b|k\rangle_{\cal A} \ .
\end{equation}
At this point, we impose the additional condition that the QCM is covariant under $SU(2)\times SU(2)$ in the computational basis (or, equivalently, under $SO(4)$ in the magic basis). This means that the QCM acts similarly in all
bases connected by local unitaries to the computational basis. 
This restriction is natural since it guarantees that all states equivalent up to local unitaries (thereby having the same entanglement) result in 
equally entangled clones.  
A sufficient condition for covariance is \cite{patrick} 
\begin{equation} \label{covariance}
|\mathcal{S} \rangle_{{\cal R},a,b,{\cal A}} = R^{\otimes 4} \;
|\mathcal{S} \rangle_{{\cal R},a,b,{\cal A}} \ ,
\end{equation}
where $R$ is any real rotation matrix in $SO(4)$. 
This requirement implies 
that $s_{ijkl}$ is an invariant tensor of rank four, that is,
$s_{ijkl}=R_{ii'}R_{jj'}R_{kk'}R_{ll'}\, s_{i'j'k'l'}$.
A main simplification here results from the fact that
the generic form of such a tensor is
\begin{equation}  \label{generic}
s_{ijkl}=A\; \delta_{il}\delta_{jk} + B\; \delta_{jl}\delta_{ik} + C\; \delta_{kl}\delta_{ij} \ ,
\end{equation}
with the normalization condition on (\ref{cloningtrans}) 
imposing that
\begin{equation}  \label{eq18}
4(|A|^2 + |B|^2 + |C|^2) + 2\,{\rm Re}(AB^*+AC^*+BC^*) = 1 \ .
\end{equation}
\par

For a symmetric cloner, the permutation symmetry between the two clones imposes 
furthermore that $A=B$, so that we are left with a transformation depending on two parameters, $A$ and $C$. If we use the cloning fidelity as a figure 
of merit, Eqs. (\ref{cloningtrans}) and (\ref{generic}) result in
\begin{eqnarray}  \label{F}
F= \langle \Phi | \rho_{a,b}| \Phi \rangle 
= 7|A|^2+|C|^2+4\,{\rm Re}(AC^*) \ ,
\end{eqnarray}
where $\rho_{a(b)}$ denotes the reduced density matrix of clone $a(b)$.
This expression can be maximized under the normalization constraint 
Eq.~(\ref{eq18}), giving
\begin{eqnarray} \label{solution}
A=\frac{1}{3}\bigg(\frac{1}{2}+\frac{1}{\sqrt{13}}\bigg)^{1/2}, \quad 
C=\frac{A}{2}(\sqrt{13}-3) \ , 
\end{eqnarray}
with the corresponding fidelity
\begin{equation}  \label{fidelity}
F=\frac{5+\sqrt{13}}{12}\simeq 0.7171 \ .
\end{equation}
Interestingly, this maximization procedure yields a fidelity 
that saturates the upper bound derived from the no-signaling condition 
in \cite{patrick}.
We therefore conclude that the optimal covariant cloner of ME 2-qubit states 
is characterized by Eq.~(\ref{solution}) and yields the 
fidelity (\ref{fidelity}). This is slightly higher 
than the fidelity of the universal 4-dimensional cloner, namely $F=7/10$ 
\cite{Werner,Cerf}, as expected since the ME states form a subset of the 2-qubit states.
\par

\begin{figure}[!t!]
\centerline{\includegraphics[width=7.0cm]{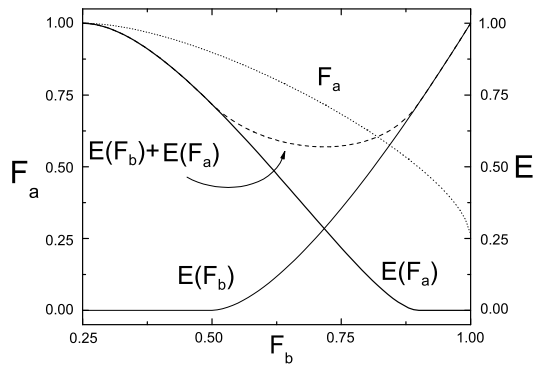}}
\caption{
Fidelity (dotted line) of clone $a$ as a function of that of clone $b$ for the optimal 2-qubit entanglement cloner. The two fidelities coincide at $F_a=F_b=(5+\sqrt{13})/12$.
Entanglement of formation of clones $a$ and $b$ (solid line)
and their sum (dashed line)
as a function of the fidelity of clone $b$. The two curves intersect at
$E_a=E_b=0.2847$.}
\label{fig1}
\end{figure}

We now generalize Eq.~(\ref{F}) to asymmetric cloning transformations 
($A\ne B$). The cloning fidelities become
\begin{eqnarray}  \label{eq17}
F_a= 4|A|^2+|B|^2+|C|^2+2\,{\rm Re}(AB^*+AC^*+BC^*) \ ,\nonumber \\
F_b= 4|B|^2+|A|^2+|C|^2+2\,{\rm Re}(AB^*+AC^*+BC^*) \ .\nonumber\\
\end{eqnarray}
Assuming $A$, $B$, and $C$ to be real, we eliminate $A$ and $C$ 
from Eqs.~(\ref{eq18}) and (\ref{eq17}), to get 
\begin{eqnarray}  \label{eq_asymmetric}
F_a(B,F_b)=-3B^2+\frac{F_b+1}{2}+\frac{\sqrt{-3B^2+F_b}-B}{2}\nonumber \\
\times  \sqrt{18B^2+18B\sqrt{-3B^2+F_b}-15F_b+6} \ .
\end{eqnarray}
We then maximize $F_a$ over $B$ as a function of $F_b$.
Figure~\ref{fig1} displays the resulting $F_a$ 
as a function of $F_b$.
The mid-point value of this curve lies at $F_a=F_b=F$, as expected.
We also confirm that the fidelity of a clone is 
equal to one when the other one is in a completely mixed state, 
i.e., when its fidelity equals 1/4. 
\par

We have numerically confirmed the optimality of this class of
asymmetric cloners with the use of a technique
based on semidefinite programming \cite{Fiurasek01,Fiurasek03}. 
The cloning transformation is 
a linear trace-preserving completely positive (CP) map that can be represented 
by a positive semidefinite operator $S$ on the tensor-product 
space of the input and output states. The cloning fidelities
can be expressed as $F_{a(b)}=\mathrm{Tr}[S R_{a(b)}]$,
with appropriately defined $R_{a(b)}\geq 0$ \cite{Fiurasek03}. The optimal asymmetric cloner can be 
obtained by maximizing $F=pF_a+(1-p)F_b$, where $p\in[0,1]$ is the
asymmetry parameter. The resulting fidelities coincide with those obtained
from Eq.~(\ref{eq_asymmetric}) up to the machine precision. 
\par

Let us now investigate the entanglement properties of this cloning 
transformation and show that it is also optimal with respect to
our original goal, namely cloning the amount of entanglement. Let
us start by checking that it preserves separability.
For the ansatz (\ref{Psidefinition}), we have $S=\mathrm{Tr}_{\cal{A}}(|\mathcal{S}\rangle\langle\mathcal{S}|)$,
and the CP map that describes
the relationship between the input and the clone $a$ ($b$)
can be characterized by $S_{a(b)}=\mathrm{Tr}_{b(a)}[S]\ge 0$.
Since the positive partial transpose (PPT) criterion is a
necessary and sufficient separability condition for a qubit pair, 
a sufficient condition for these two maps to preserve
separability is that $S_a$ and $S_b$ represent PPT operations \cite{Rains01}.
The map $S_{a(b)}$ is PPT if
$S_{a(b)}^{T_{1,1'}}\geq 0$, where $T_{1,1'}$ denotes partial transposition
with respect to the first qubit of the original and the clone $a(b)$. 
An explicit analytical calculation shows that if
$s_{ijkl}$ is an invariant rank-4 tensor (\ref{generic}), then
$S_{a(b)}^{T_{1,1'}}=S_{a(b)}\geq 0$, hence a covariant cloner necessarily 
preserves separability.
\par

It is therefore natural to maximize the output entanglement for the other extreme case, namely when the original qubit pair 
is maximally entangled. 
The amount of entanglement left in the clones 
will be measured here by the entanglement of formation $E$,
which can be evaluated by using the extended 
definition of the concurrence ${\cal C}$ for mixtures \cite{HW}.
The entanglement of formation of an arbitrary 2-qubit state $\rho$ is 
given by $E(\rho)=E({\cal C}(\rho))$, where 
${\cal C}(\rho)= \max (0,\lambda_1-\lambda_2-\lambda_3-\lambda_4)$ and the $\lambda_i$'s 
are the eigenvalues, in decreasing order, of the Hermitian matrix 
${\tilde \rho} \equiv \sqrt{\sqrt{\rho}\rho^*\sqrt{\rho}}$. Here $\rho^*$ denotes the complex conjugate of $\rho$ when expressed in the magic basis \cite{HW}.
For a generic covariant cloner, Eq.~(\ref{generic}), the reduced density matrices of the clones of an input ME state $|e_i\rangle$ can always
be written as
\begin{eqnarray}
\rho_{a,b} = F_{a,b} \, |e_i\rangle \langle e_i | + \frac{1-F_{a,b}}{3} \sum_{j\not= i}|e_j \rangle \langle e_j |  \ ,
\end{eqnarray}
so the clones are left in a mixture of (generalized) Bell states, or more precisely in a Werner state.
This is consistent with the fact that $\rho_{a,b}$ are 
real density matrices in the magic basis because
the input state is real \cite{HW}. Hence,
${\tilde \rho_{a,b}}=\rho_{a,b}$ so the concurrence of the clones simply reduces to ${\cal C}_{a,b}=\max(0,2F_{a,b}-1)$. Therefore, for a covariant cloner, maximizing
$E$ reduces to maximizing $F$. Consequently, the cloner
characterized by Eq.~(\ref{solution}) is the entanglement cloner we were
looking for, provided that covariance is taken for granted.
\par

The corresponding entanglement of formation of the clones,
$E_a$ and $E_b$, is shown in Fig.~\ref{fig1}
for different values of $F_b$. As expected,
the entanglement of formation of clone $b$ vanishes for $F_b \le 1/2$. Conversely, the entanglement of formation of clone $a$ vanishes when $F_b \ge 0.8984$ (that is, when $F_a \le 1/2$). 
Note that the entanglement of formation of a clone is equal to one only when its fidelity is one, thus confirming that a fully asymmetric cloner (a trivial cloner which outputs the original and a random clone) is the only solution if we want to fully conserve the original entanglement of 1 ebit.
Finally, note that the sum of the entanglement of formation of the two clones (also shown in Fig.~\ref{fig1}) never exceeds one, meaning that the entanglement cloner does not create more entanglement than that contained in the original ME state.
\par

For the symmetric cloner, the entanglement of formation of both clones 
is equal to $E_a=E_b=0.2847$~bits.
The optimality of this result has been verified using numerical optimization where the structure of the cloning transformation 
was based on the parametrization proposed in \cite{Cerf} and the maximized
quantity was the concurrence instead of the fidelity. Up to irrelevant
local unitaries (which decrease $F$ while keeping ${\cal C}$ and $E$ constant),
we recovered the same cloning transformation. This strongly 
suggests that restricting ourselves to covariant QCMs is justified,
so the cloner of ME states also optimally clones the amount of entanglement.
\par

\begin{figure}[!t!]
\centerline{\includegraphics[width=7.0cm]{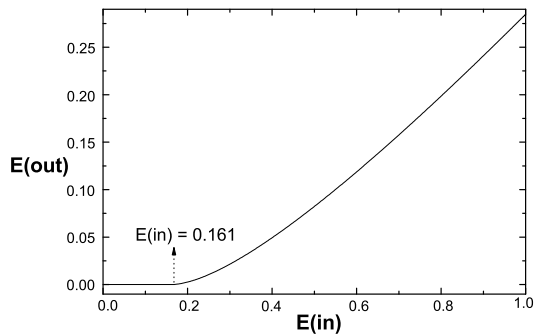}}
\caption{
Entanglement of the clones versus the entanglement of the input state 
for the symmetric cloner.}
\label{fig2}
\end{figure}

Finally, the properties of our entanglement cloner can be analyzed
in the intermediate case of non-maximally entangled input states.
In Figure \ref{fig2}, we plot the entanglement of formation
of the clones as a function of that of the original state $\alpha|00\rangle
+\sqrt{1-\alpha^2}|11\rangle$ ($0\le\alpha\le 1$) for the symmetric cloner. 
No entanglement is cloned below a critical value  
$E_{\rm in}=0.161$~bits. Then, the output entanglement increases monotonically
to reach its maximum $E_{\rm out}=0.2847$~bits at $E_{\rm in}=1$~bit.
\par

In conclusion, we have shown that the quantum entanglement of an
unknown ME qubit pair cannot be perfectly cloned if, at the same time, product states are required to be cloned into unentangled qubit pairs. In other words,
a separability-preserving QCM cannot perfectly duplicate the entanglement of  
the set of ME states. Only imperfect QCMs do exist.
As a first step, we have constructed an optimal symmetric entanglement cloner which is universal over the set of ME states, and whose fidelity saturates the no-signaling upper bound \cite{patrick}.
This cloner yields imperfect clones with
0.285~ebits if the original qubit pair contains 1~ebit, while unentangled 
pairs are cloned into separable states. 
The distribution of entanglement among the clones has also 
been investigated using an asymmetric cloner.
Similarly to the situation that prevails when cloning quantum states, 
this no-go theorem for entanglement cloning might be exploited in order to imagine new quantum key distribution schemes. For example, 
one could imagine a protocol where the eavesdropper is only able to
apply a {\em local} cloning on a ME state instead of the above global cloning. 
One can check that applying the optimal universal qubit cloner \cite{BH}
locally on each qubit of a ME state reduces
the fidelity of the clones to $7/12$. Since the reduced density
matrix of the clones is again a Werner state, we obtain 
${\cal C} = 1/6$ leading to only about 0.060~ebits per clone.
Thus, the fact that independent local operations on each qubit 
are less efficient than joint operations could be used to give an advantage
to the authorized parties.
\par

We thank T. Durt and S. Iblisdir for helpful discussions.
We acknowledge funding from the Communaut\'e Fran\c caise de
Belgique under grant ARC 00/05-251, from the IUAP programme of the Belgian
government under grant V-18, and from the EU under project CHIC
(IST-2001-32150). JF also acknowledges funding
from the  grant LN00A015  of the Czech Ministry of Education.

\end{document}